\documentclass{article}    

\usepackage{graphicx}

\title{\textbf{Conscious Pulse I: \mbox{The rules of engagement}}}  
\author{Richard Mould\footnote{Department of Physics and Astronomy, State University of New York, Stony Brook,
\mbox{New York} 11794-3800; http://nuclear.physics.sunysb.edu/ \~{}mould}}  
\date{}    
   
\begin{document}             

\maketitle              

\begin{abstract}

      	This paper elaborates on four previously proposed rules of engagement between conscious states and
physiological states.  A new rule is proposed that applies to a continuous model of conscious brain states that cannot
precisely resolve eigenvalues.  If two apparatus states are in superposition, and if their eigenvalues are so close
together that they cannot be consciously resolved on this model, then it is shown that observation will not generally
reduce the superposition to just one of its member eigenstates.  In general, the observation of a quantum mechanical
superposition results in another superposition.   

\end{abstract}

\section*{Introduction}

	The author has proposed four rules that describe the relationship between conscious states of the brain and quantum
physiology.   In one paper, the rules are successfully applied to a typical quantum mechanical interaction between a
particle and a detector \cite{RM1}; and in another paper, they are successfully applied to two different versions of
the Schr\"{o}dinger cat experiment \cite{RM2}.  In this paper, the third rule is expanded to cover the case of
continuous brain states; and in a future paper, a final rule will be added that also applies to this continuous
\mbox{case \cite{RM3}}.  

The first rule of the previous papers introduce quantum mechanical probability through the positive flow of
\emph{probability current} $J$, which is equal to the time rate of change of square modulus.  Probability is not
otherwise defined in this treatment.   

\vspace{0.5 cm}

\textbf{Rule (1)}\emph{ For any subsystem of n components in an isolated system with a square modulus equal to s, the
probability per unit time of a stochastic choice of one of those components at time t is given by $(\Sigma_nJ_n)/s$, where the net
probability current $J_n$ going into the $n^{th}$ component at that time is positive.   }

The \emph{ready brain state} referred to in rule (2) is not conscious by definition, but it is physiologically
capable of becoming conscious if it is stochastically chosen.  

\textbf{Rule (2)}\emph{If the Hamiltonian gives rise to new components that are not classically continuous with the old
components or with each other, then all active brain states that are included in the new components will be ready brain
states.}[Active brain states are either conscious  or ready states.].

The third rule describes a state reduction like Penrose's process \textbf{R}.  It is understood to provide a
new boundary condition.  

\textbf{Rule (3)}: \emph{If a component that is entangled with a ready brain state B is stochastically chosen, then B will become
conscious, and all other components will be immediately reduced to zero.}

The fourth rule is added to prevent certain anomalies from occurring as a result of the first three rules by
themselves.  

\textbf{Rule (4)}\emph{A transition between two components is forbidden if each is an entanglement containing a
ready brain state of the same observer}

\vspace{0.5 cm}

As was our practice in the previous papers, a conscious brain state will be represented by an underlined
$\underline{B}$, and a ready brain state $B$ will appear without an underline.  In this paper, the different brain
types
 $\underline{B}_k(\alpha)$ and $B_k(\alpha)$ for a particular state variable $k$  are given as a function of brain
variables $\alpha$.  For both types we require.  
\begin{equation}
\int \!\!d\alpha\,\underline{B}_r(\alpha)^*\underline{B}_s(\alpha) = \delta(r-s)       \hspace{.5cm}\mbox{and
}\hspace{.3cm}    
\int \!\!d\alpha\,{B}_r(\alpha)^*{B}_s(\alpha) = \delta(r-s)
\end{equation}

\section*{A Conscious Brain Pulse - Rule (3a)}

		I assume that there is a limit to how sharply a conscious experience can be defined.  It is unphysical to imagine
that a precisely defined physiological state can support a knife-edge slice of consciousness.  That is, a
physiological state $\underline{B}_k$ with exact eigenvalues cannot be expected to support ``recognizable"
consciousness without involving other states in its immediate neighborhood.  Any real conscious experience therefore
engages a group of neighboring states that will hereafter be designated by the symbol $\{\underline{B}_k\}$, where
the brackets around $\underline{B}_k$ specify a group of states with $\underline{B}_k$ at its center.  I call this
collection of states a \emph{conscious brain pulse}, or just a \emph{conscious pulse}.  It is given by
\begin{eqnarray}
\{\underline{B}_k\} \!\!\!&=& \!\!\!\int \!\!du\,F_k(u)\underline{B}_u \\
 where \!\!\!&&\!\!\!\int \!\!du\,F_k(u)^*F_k(u) = 1\nonumber
\end{eqnarray}

The states in $\{B_k\}$ are not a statistical mixture because $F_k(u)$ represents the coefficients of a continuous
superposition of quantum mechanical states $\underline{B}_u$.  Although these have macroscopic dimension, 
they cannot display local interference effects because of environmental decoherence as
explained in \mbox{ref.\ 1}. 

The \emph{conscious experience} that is associated with a conscious brain pulse will result from the collective effect
of all the conscious states in the pulse neighborhood, where the width of this pulse reflects a limit on the ability of
the brain to resolve the experience.  

	A ready brain state is not conscious; nonetheless, it will generally exist as a similar collection of states $\{B_k\}
=  \int \!\!du\,F_k(u)B_u$ that will be called a \emph{ready brain pulse}\footnote{A ready pulse generally evolves from
a conscious pulse, and will therefore take on the functional form of that pulse.  However, a ready state may also evolve under
Schr\"{o}dinger from a single `unconscious' state, in which case it will be a \emph{single} ready state.}.   If current flows into a
component containing a pulse of ready brain states, and if one of those states given by $B_{sc}$ is stochastically chosen from
the pulse according to rule (1), then it will become conscious according to rule (3).  What happens after that is
determined by the properties of the brain.  Specifically, the final result of a stochastic selection is not just the
single state $\underline{B}_{sc}$, but the entire conscious pulse $\{\underline{B}_{sc}\}$.   After the pulse is
formed, the special status of $\underline{B}_{sc}$ is lost, except as it identifies the maximum of the resulting
pulse.  It follows from the above definitions that the conscious and ready brain pulses are themselves normalized.  
\begin{equation}
\int \!\!d\alpha\,{\{\underline{B}}_k\}^*\{\underline{B}_k\} = 1       \hspace{.4cm}\mbox{and}\hspace{.4cm}    
\int \!\!d\alpha\,{\{B}_k\}^*\{B_k\} = 1
\end{equation}

We will now supplement rule (3) by adding rule (3a).  This describes what happens to a stochastically
chosen ready brain state in the present model.  The rule (3) conversion to a conscious state, and the reduction of all
other states to zero is assumed to take place in a single instant of time.  After that, the brain's Hamiltonian
will form a conscious pulse at a more leisurely physiological pace.
 
\vspace{0.5 cm}
\noindent
\textbf{Rule (3a)}: \emph{The Hamiltonian of the brain will convert a chosen conscious state into a conscious pulse
whose width reflects the ability of the brain to resolve the conscious experience.}  
\vspace{0.5 cm}

	Classically, a conscious experience is prompted by an external stimulus that may be very sharply defined; and yet,
there is a limit to how sharply it can be experienced by the viewer.  We classically deal with this by assuming that
such an incoming `sharp' signal is spread out by physiological constraints contained in the Hamiltonian.  In the same
way, rule (3a) claims that a single stochastically chosen conscious state is converted by the brain into a conscious
pulse, thereby providing a space in the brain for a full conscious experience.  

	When a sharply defined stochastically chosen state dissolves into a broadly defined pulse, discharge current will
flow from it to its immediate neighbors.  In the process a normalized single state $\underline{B}_k$ 
becomes a normalized pulse $\{\underline{B}_k\}$, thereby conserving current.

\section*{An Interaction}

			   	In an interaction like the one described in the previous paper, a conscious brain state is initially correlated
with an apparatus state $A_1(t)$, where the system evolves under Schr\"{o}dinger into a ready brain state that is
correlated with another apparatus state $A_2(t)$.  Rule (2) requires the evolution of  ready brain states only.  Let $A_1(t)$ be
normalized to 1.0 at $t_0$ = 0 and decrease in time, and let $A_2(t)$ be zero at $t_0$ and increase in time.  We now amend the
previous description given in refs.\ 1 and 2 to refer to pulses rather than states.  

Let the initial state of the system be given by $A_1(t)\{\underline{B}_1\}$, where $\{\underline{B}_1\}$ is the
initial conscious pulse of the observer who is aware of the apparatus state $A_1$; and let every individual brain
state in this pulse evolve under Schrš\"{o}dinger into a corresponding  `ready' brain state.  The emerging
component in eq.\ 4 is then $A_2(t)\{B_2\}$, and the system prior to a stochastic choice at $t_{sc}$ is
\begin{equation}
\Phi(t_{sc} > t \ge t_0) = A_1(t)\{\underline{B}_1\} + A_2(t)\{B_2\}         
\end{equation}
where the entanglement $A_1(t)\{B_2\}$ is initially equal to zero.\footnote{As in previous papers, the pre-interaction
apparatus states $A_1$ or $A_2$ are different than the entangled apparatus states in eq.\ 4 because the latter include
the ``low level" physiology of the observer.  In this case, the entangled apparatus states must fan-out at the
physiology end into a superposition that connects with each component of the brain pulses.}

At the time of stochastic choice, a single ready state $B_{sc}$ in $\{B_2\}$ is selected and made conscious, with all
other components going to zero as per rule (3). 
\begin{displaymath}
\Phi(t_{sc})= A_2(t_{sc})F_2(sc)\underline{B}_{sc}
\end{displaymath}
Rule (3a) requires that the single state $\underline{B}_{sc}$ subsequently becomes a pulse in physiological time. 
\begin{equation}
\Phi(t > t_{sc}) = A_2(t_{sc})F_2(sc)\{\underline{B}_{sc}\}
\end{equation}
The probability that the state (sc) in $\{B_2\}$ is stochastically chosen can be found from the second component of eq.\ 4 by using
the Born rule. 
\begin{eqnarray}
P(sc) \!\!\!&=& \!\!\!(1/\!s)\!\!\int \!\!dx\!\!\int \!\!d\alpha\,
A_2^*A_2F_2(sc)^*F_2(sc)\{\underline{B}_{sc}\}^*\{\underline{B}_{sc}\}\nonumber\\
 &=& \!\!\!(1/\!s)F_2(sc)^*F_2(sc)\!\!\int \!\!dx\, A_2^*A_2 \nonumber
\end{eqnarray}
where $x$ refers to the apparatus variables, and $s$ is the square modulus of the first component in eq.\ 4.  The
total probability of a stochastic hit in the ready pulse is then found by integrating over $d(sc)$.
\begin{displaymath}
P = (1/\!s) \!\!\int \!\!d(sc)\,F_2(sc)^*F_2(sc)\!\!\int \!\!dx\,A_2^*A_2 = (1/\!s)\!\!\int \!\!dx \,A_2^*A_2
\end{displaymath}
where $A_2^*A_2$ is the square modulus when the interaction  is complete.

The central state $\underline{B}_{sc}$ of the conscious pulse in eq.\ 5 is included in the original ready pulse
$\{B_2\}$, but it is not necessarily the central state $B_2$.  Therefore, the stochastically chosen state cannot be
exactly determined by the Hamiltonian, due to the inability of the brain to fully resolve the ready brain states that
are candidates for stochastic selection.   As in previous cases, the reduction in eq.\ 5 is not normalized.  This does
not affect probability calculations so long as \mbox{rule (1)} is faithfully followed.

\section*{Unresolvable Observation}

			   	Let the system be a stationary superposition of apparatus states $A_1$ and $A_2$ at time $t_0$.  
\begin{equation}
\Phi(t_0) = (A_1 + A_2)\{\underline{X}\}
\end{equation}
where $\{\underline{X}\}$ is an unknown conscious state of an observer who has not yet interacted with the
apparatus.  At time $t_{ob}$ the observer looks at the apparatus, and the system becomes  
\begin{eqnarray}
\Phi(t \ge t_{ob} > t_0) \!\!\!&=& \!\!\![A_1(t) + A_2(t)]\{X\}\nonumber\\
 &+& \!\!\!A_1'(t)\{B_1\} + A_2'(t)\{B_2\}\nonumber
\end{eqnarray}
following rule (2).  The primed components are zero at $t_0$.  Substituting eq.\ 2
\begin{eqnarray}
\Phi(t \ge t_{ob} > t_0) \!\!\!&=& \!\!\![A_1(t) + A_2(t)]\{X\}\nonumber\\
 &+& \!\!\!\int \!\!du\,[A_1'(t)F_1(u) + A_2'(t)F_2(u)]B_u\nonumber
\end{eqnarray}
where the primed components in the second row increase and the unprimed components in the first row go to zero in
physiological time.  As current flows from the first to the second row, there is certain to be a stochastic hit on one
of the ready brain states according to rule (1).   Looking at the system at the moment rule (3) applies, but before
rule (3a) can take effect, we find the reduction
\begin{displaymath}
\Phi(t = t_{sc} > t_{ob}) = [A_1'(t_{sc})F_1(sc) + A_2'(t_{sc})F_2(sc)]\underline{B}_{sc}
\end{displaymath}
Rule (3a) now requires that the state $\underline{B}_{sc}$ dissolve into a pulse.  
\begin{equation}
\Phi(t > t_{sc}) = [A_1(t_{sc})F_1(sc) + A_2(t_{sc})F_2(sc)]\{\underline{B}_{sc}\}
\end{equation}
where the primes on $A_1$ and $A_2$ are dropped.

If the functions $F_1(sc)$  and $F_2(sc)$ do not overlap, then a stochastic choice will pick out a state in either
$F_1$ or $F_2$.  However, it is possible that the pulses do overlap as
shown in fig.\ 1, and that the stochastic choice picks out a state in the overlap.  In that case, the amplitude
of the chosen pulse will be the entire bracketed coefficient of the pulse that appears in eq.\ 7.   

\vspace{.2cm}
\begin{figure}[h]
\centering
\includegraphics[scale=0.6]{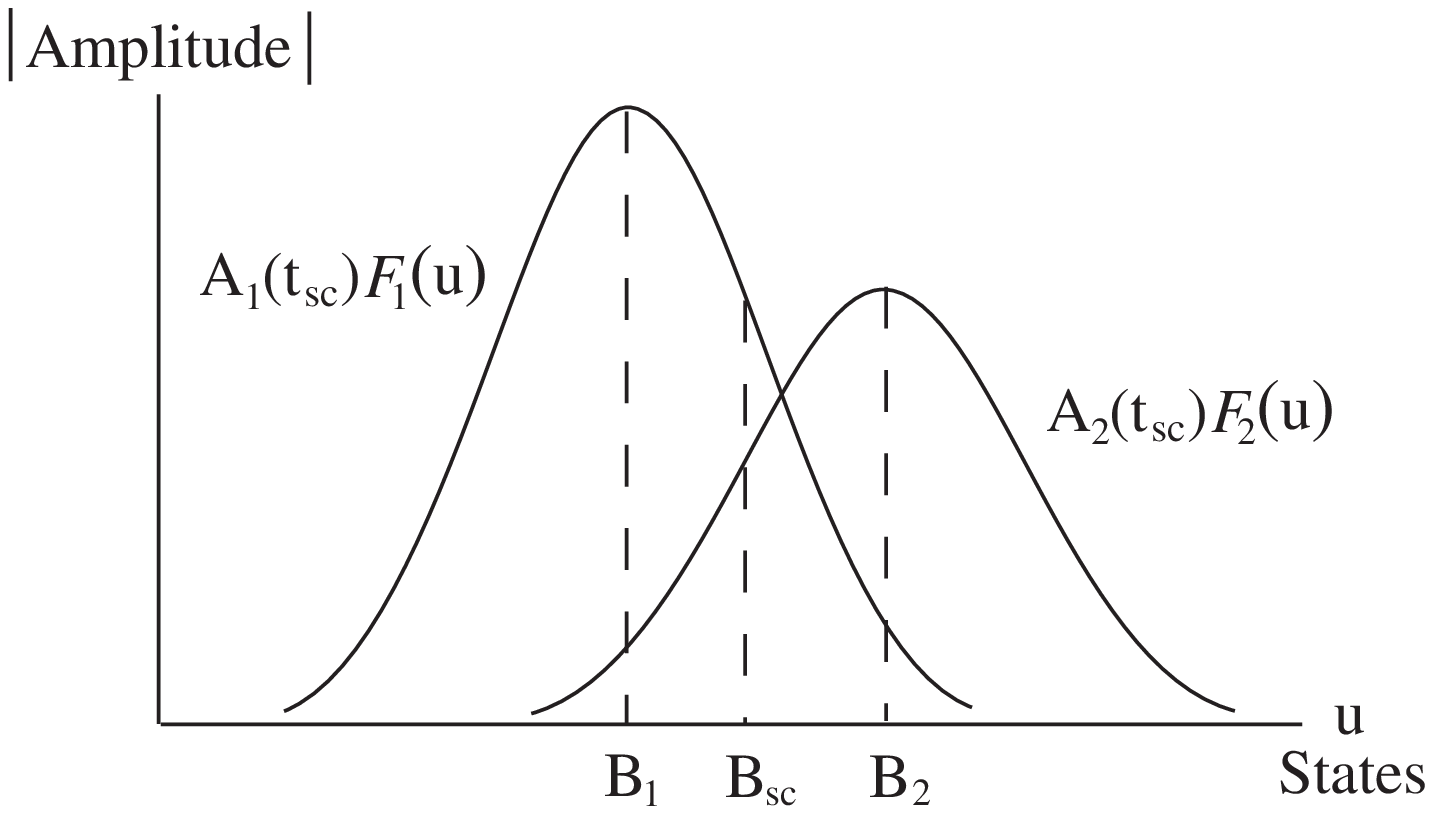}
\center{Figure 1}
\end{figure}

Evidently the initial apparatus superposition in eq.\ 6 is replaced by a different superposition in eq.\ 7.  The
observer fails to reduce the initial superposition to just one of the two eigenstates, because he cannot fully resolve
the two eigenvalues.

The experimental meaning of the superposition in eq.\ 7 can be clarified by disabling one of the apparatus states, say
$A_1$, and noting the probability that $A_2$ continues to be observed.  For example, imagine that the observable
associated with $A_1$ is a spot of light appearing on a screen, and the observable associated with $A_2$ is another
spot of light that is so close to the first that it cannot be fully resolved by the observer.  To decide if he is
looking at the first or the second spot following a stochastic choice, the observer turns off the first source of
light, and notes that the spot does or does not remain.   When that is done at \mbox{time $t_{o\!f\!f}$}, \mbox{eq.\
7} becomes   
\begin{displaymath}
\Phi(t \ge t_{o\!f\!f} > t_{sc}) = A_2(t_{sc})F_2(sc)\{\underline{B}_{sc}\}
\end{displaymath}
The probability that the spot is observed in the second apparatus state can be found by integrating the square modulus of
this expression and making use of the Born rule.  
\begin{displaymath}
P_2^{(sc)} (t> t_{o\!f\!f}) = (1/\!s)\!\!\int \!\!dx\!\!\int \!\!d\alpha \,\Phi^*\Phi = (1/\!s)\!\!\int \!\!dx\,
A_2^*A_2F_2(sc)^*F_2(sc)
\end{displaymath}
where $s$ is the square modulus of eq.\ 6.

If the experiment is performed many times, then summing over all the possible stochastic choices, the probability of
observing the second apparatus eigenvalue will be
\begin{eqnarray}
P_2 (t> t_{o\!f\!f}) \!\!\!&=& \!\!\!\int \!\!d(sc)\,P_2^{(sc)} (t> t_{o\!f\!f})\nonumber\\
 &=& \!\!\!(1/\!s)\!\!\int \!\!dx\,A_2^*A_2\!\!\int \!\!d(sc)\,F_2(sc)^*F_2(sc)\nonumber\\
&=& \!\!\!(1/\!s)\!\!\int \!\!dx\,A_2^*A_2 \nonumber
\end{eqnarray}
This is the same result that one would expect if the states $\{B_1\}$ and $\{B_2\}$ were completely resolvable.   

	It should be noted that if the observer becomes disengaged from the apparatus at some time $t_{dis}$ after the
stochastic hit in eq.\ 7, the system would become
\begin{equation}
\Phi(t \ge t_{dis}> t_{sc}) = [A_1(t_{sc})F_1(sc) + A_2(t_{sc})F_2(sc)]\{X\}
\end{equation}
where $\{X\}$ is the disengaged state that evolves from $\{\underline{B}_{sc}\}$ in physiological time.  This
expression makes the independence of the observer and the system more apparent.  The effect of the observation has
therefore been to change the system from the initial apparatus superposition $(A_1 + A_2)$ in eq.\ 6 to the
superposition $A_1(t_{sc})F_1(sc) + A_2(t_{sc})F_2(sc)$ in eq.\ 8.  The observation brings about a state reduction,
but it does not reduce the state to either $A_1$ or $A_2$ as would normally be expected.  As previously stated, this is
because the observer cannot clearly resolve the two possibilities, so he cannot clearly reduce the system to one or
the other eigenstate.  

The probability of the final state of the system in eq.\ 7 is found by integrating the
variables $dx$, $d\alpha$, and  $d(sc)$ over the entire time of the physiological interaction leading to eq. 7.
\begin{eqnarray}
P_2 (t> t_{sc}) \!\!\!&=&\!\!\!(1/\!s)\!\!\int \!\!dx\!\!\int \!\!d\alpha\!\!\int \!\!d(sc)\,[A_1F_1(sc) +
A_2F_2(sc)]^*\nonumber\\ & & \hspace{2.74cm}
\times\,[A_1F_1(sc) +,A_2F_2(sc)]\{\underline{B}_{sc}\}^*\{\underline{B}_{sc}\}  
\nonumber\\ &=& \!\!\!(1/\!s)\!\!\int \!\!dx \,[A_1^*A_1 + A_2^*A_2 ] \nonumber
\end{eqnarray}
which is the same as the probability of the initial state in eq.\ 6. 

If the unknown state $\{X\}$ in eq.\ 6 is a single unconscious state, then the resulting ready brain states that
engage the apparatus will also be single states $B_1$ and $B_2$.  In that case, it will always be possible for the
reduction to  make an unambiguous choice between $B_1$ and $B_2$.  This does not mean that the observer will be able
to psychologically resolve the two, but only that the rule (3) reduction will not lead to a superposition in
these circumstances.

\section*{Pulse Drift}

			Rule (2) requires that all newly emerging and discrete active brain states are ready states.  Clearly, the states
within a conscious pulse are intended to be psychologically indistinguishable from one another; however,
distinguishability  or discreteness in the sense of rule (2) will be given a more narrow meaning.  If the conscious pulse
$\{\underline{B}_k\}$ is said to include the \emph{immediate} neighborhood of $\underline{B}_k$ (i.e., those states
that are psychologically indistinguishable from $\underline{B}_k$), then I will say that only the \emph{most immediate}
neighbors of $\underline{B}_k$ are the ones that are exempt from rule (2), and are thereby  directly influenced by
$\underline{B}_k$ .  Only these states are pulled directly into existence by $\underline{B}_k$ during pulse formation, and they
will have a lesser amplitude than  $\underline{B}_k$.  They, in turn, will pull their most immediate neighbors into the pulse, again
with lesser amplitude.  In this way, the entire pulse is drawn into being around the initial central state $\underline{B}_k$.  

This means that the pulse does not have a definite edge.  However, there is still a decisive limit to the influence of
each state within the pulse, beyond which rule (2) applies to interactions involving that particular state.   

With this understanding, there is nothing in the rules that would prevent a conscious pulse from drifting
continuously about the brain, moving  over a wide range of brain states without the necessity of hopping
stochastically from one place to another.  As a  pulse of this kind drifts forward, the conscious states in its
leading edge will gain amplitude, and those in its trailing edge will lose amplitude, without engaging ready brain
states as required by \mbox{rule (2)}.   

	Now consider what will happen when the conscious pulse drifts continuously over the brain in this way, while at the
same time giving rise to a ready brain pulse as in eq.\ 4.  A ready brain pulse cannot move like an ordinary pulse. 
Its trailing edge cannot feed current to its leading edge because of rule (4), so the amplitude of a single component
of the ready pulse can only increase by virtue of current coming from the conscious pulse\footnote{This is another
example of how rule (4) prevents an anomalous increase in probability.  Trailing edge current flowing into the leading
edge would otherwise cause extraneous \mbox{rule (3)} reductions. Other examples are in refs.\ 1 and 2.}.  The moment
that current stops for any reason, the ready component will become a stationary ``phantom" component that serves no
further purpose\footnote{The properties of a phantom component are defined in ref.\ 1.}.  It will not follow the
motion of the conscious pulse.  So instead of there being a moving ready brain pulse that parallels the motion of a
conscious pulse of decreasing amplitude, there will be a trail of ready states that become phantoms the moment they settle down to
a constant amplitude.

\section*{Intensity of a Conscious Experience}

				In classical physics, intensity is proportional to square amplitude; whereas in standard quantum mechanics,
intensity is implicit in the definition of a state rather than in its amplitude.  That's because the square modulus in
a quantum mechanical state refers \emph{only} to probability in a standard quantum mechanical treatment; and in the
present treatment it doesn't even do that.  So in the quantum case, a non-zero conscious state is
always \emph{fully} conscious, independent of its amplitude.  This is why we require that a stochastically chosen
conscious state $\underline{B}_k$ is normalized to 1.0.  It will be either on or off.  It can have no intermediate
value.  This is also why a conscious pulse $\{\underline{B}_k\}$ is normalized to 1.0.  It too can have no
intermediate value.  Of course the component in which the state or pulse appears can have intermediate values, but the
on-off nature of consciousness is represented here by the normalization of a state or a pulse, not by a component.  

 The quality of consciousness (including intensity) is governed in every case by the Hamiltonian.  So the intensity
of a psychological experience that is associated with a conscious pulse $\{\underline{B}_k\}$ is a function of the 
definition of the states that are involved.  It is one thing if a state constitutes an experience on a \mbox{sun-lit}
landscape, and another if it is an experience in a darkened basement.  In either case, the Hamiltonian of the state
will assign a lesser intensity to the neighborhood states surrounding the central state.  This means that the
intensity of the observer's experience will fade out at the edge of a conscious pulse.  We represent this modulation
of intensity by the function $F_k(u)$ in eq.\ 2.  

If we quantify the ``intrapulse" intensity $I$ by saying that it equals 1.0 for each conscious pulse (corresponding to
each pulse  being fully conscious), then \mbox{$dI = F_k(u)^*F_k(u)du$} will be the \emph{relative intensity}
of the differential range of states in the vicinity of $\underline{B}_k$.  The square modulus of $\underline{B}_k$ does not have a
formal interpretation in this treatment, but its  intensity relative to other states within a pulse can certainly be represented in this
way.

\section*{Fading in and out}

				The question then is: does a fully conscious experience arise discontinuously when a conscious pulse comes into
being?  And conversely, is the experience turned off discontinuously as a conscious pulse is reduced to zero?  The
rules are flexible enough to allow the Hamiltonian to introduce or withdraw consciousness continuously over finite
intervals of time.

\vspace{.2cm}
\begin{figure}[h]
\centering
\includegraphics[scale=0.7]{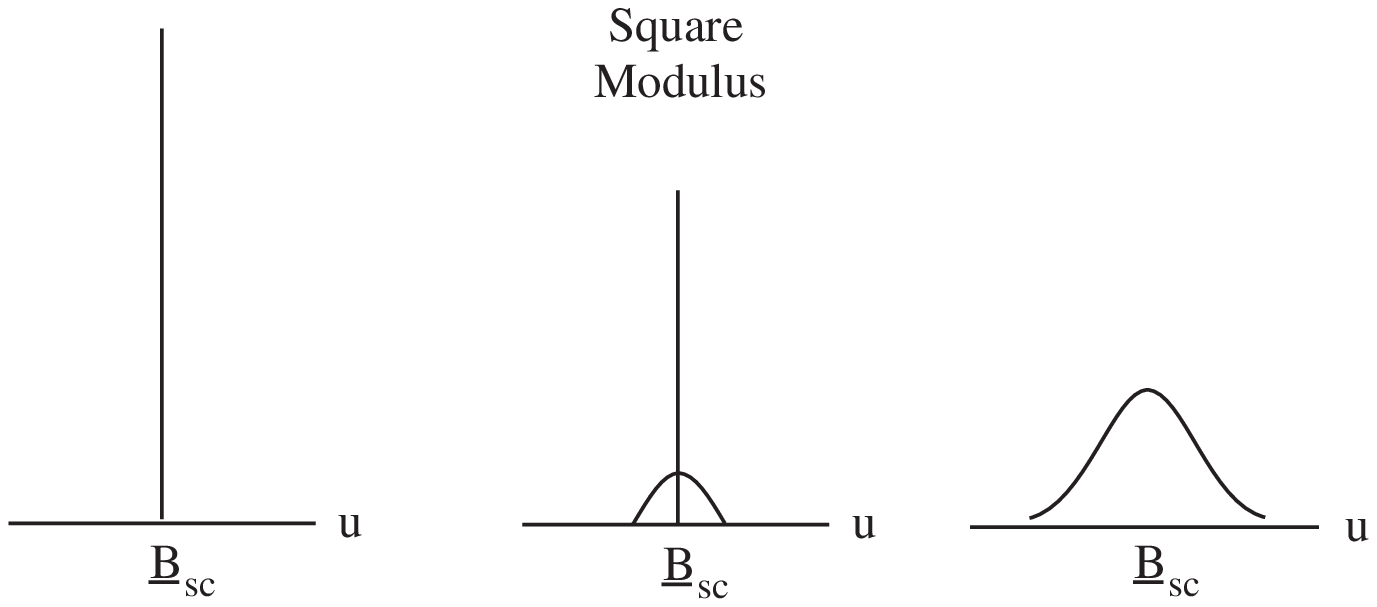}
\center{Figure 2}
\end{figure}

 The first stage in fig.\ 2 shows the stochastically chosen state the moment it is created.  The Hamiltonian reduces
its amplitude in the second stage, giving rise to a pulse that only involves its ``most immediate" neighbors.  In the
third stage, the initial state is completely absorbed into the pulse, and the width of the pulse has expanded to a
degree that allows a full conscious experience.  Although the initial state is technically conscious, it is too narrow
to support a recognizable psychological experience.  The number of states involved in the second stage of fig.\ 2 will
support some degree of the full experience, but only the third stage supports the full experience.  This sequence
allows the observer to become gradually aware of the pulse on a time scale that is governed by the Hamiltonian.  At
the same time, it does not violate the on-off principle that is represented in the normalization of the
state-plus-pulse. 

The converse cannot be true in the same way.  Rule (3) requires that a conscious state will go immediately to zero if
there is a stochastic choice of another state; and this suggests that there can be no gradual phasing out of a
conscious experience.  However, there may be another mechanism that will come to the rescue.  The Hamiltonian might
provide for the existence of an ``after glow" of any terminated conscious experience.  This could occur through
another interaction that is in parallel with the primary interaction; and it might well be related to the interaction
that puts any conscious experience into short-term memory.  If that is true, then the Hamiltonian would control the
extent to which the observer fades in or out of consciousness, and that is certainly the desired result.

\end{document}